\documentclass[12pt]{iopart}
\usepackage{graphics}
\usepackage{graphicx}
\usepackage{color}
\pdfoutput=1
\begin{document}

\title[Modulation transfer spectroscopy in atomic rubidium]{Modulation transfer spectroscopy in atomic rubidium}

\author{D. J. McCarron, S. A. King and S. L. Cornish}

\address{Department of Physics, Durham University, Durham, DH1 3LE, UK}
\ead{s.l.cornish@durham.ac.uk}

\begin{abstract}
We report modulation transfer spectroscopy on the $D2$ transitions in $^{85}$Rb and $^{87}$Rb using a simple home-built electro-optic modulator (EOM). We show that both the gradient and amplitude of modulation transfer spectroscopy signals, for the  $^{87}$Rb $F=2 \rightarrow F^{'}=3$ and the $^{85}$Rb $F=3 \rightarrow F^{'}=4$ transitions, can be significantly enhanced by expanding the beams, improving the signals for laser frequency stabilization. The signal gradient for these transitions is increased by a factor of 3 and the peak to peak amplitude was increased by a factor of 2. The modulation transfer signal for the $^{85}$Rb $F=2 \rightarrow F^{'}$ transitions is also presented to highlight how this technique can generate a single, clear line for laser frequency stabilization even in cases where there are a number of closely spaced hyperfine transitions.

\end{abstract}

\maketitle

\section{Introduction}
Many atomic physics experiments, especially those in the field of
laser cooling \cite{nobel97} rely upon the active frequency
stabilization, or `locking' of a laser. Many techniques
exist to obtain a signal which can be used to regulate the frequency
of the laser (the `error signal'). These techniques include the
dichroic atomic vapour laser lock (DAVLL)
\cite{Corwin98,Fred07,Danny07}; a combination of DAVLL and
saturation absorption \cite{Harris08}; Sagnac interferometry
\cite{Jundt03}; polarization spectroscopy \cite{Wieman76,Pearman02};
frequency modulation (FM) spectroscopy \cite{Bjorklund80}; and
modulation transfer spectroscopy \cite{Shirley82,Zhang03}. Single beam techniques, such as DAVLL, have Doppler-broadened spectral features and consequently exhibit a capture range (defined as the frequency excursion the system can tolerate and still return to the desired lock-point) of several hundred MHz.  The pump-probe schemes listed above, such as modulation transfer spectroscopy, achieve sub-Doppler resolution and as a result display much steeper signal gradients and enhanced frequency discrimination, though at the expense of a more limited capture range (typically below one hundred MHz). In particular, modulation transfer spectroscopy has two clear advantages over other techniques. Firstly, the technique readily generates dispersive-like lineshapes which sit on a flat, zero background. Consequently, the zero-crossings of the modulation transfer signals are accurately centred on the corresponding atomic transitions. Secondly, the signals are dominated by the contribution from closed atomic transitions. This second feature is especially useful when the spectrum in question contains several closely spaced transitions.

In this
work we provide an experimental study of modulation transfer
lineshapes, obtained with the $D2$ transitions in $^{85}$Rb and
$^{87}$Rb, characterizing their dependence on the modulation
frequency with a view to establishing the experimental parameters that yield the optimum lineshapes for laser locking. As the frequency stability of any locked laser depends on many parameters (including the passive stability of the laser design and the performance of the servo electronics) we confine our discussion to the behavior of the amplitude and gradient of the modulation transfer signal.  The structure of the remainder of the paper is as
follows. Section 2 outlines the origin of the modulation transfer
spectroscopy signal. Section 3 describes the experimental apparatus, provides details of the methodology and presents a simple design for a home-built electro-optic modulator (EOM). Section 4 presents our experimental results, focussing on the dependence of the modulation transfer signal on the modulation frequency and beam size. In section 5 we draw our conclusions.

%

\section{Origin of the modulation transfer lineshape}\label{theory}

\begin{figure}[h!]
\centering
\includegraphics[angle=270, trim = 10mm 10mm 45mm 0mm, clip, scale=0.79]{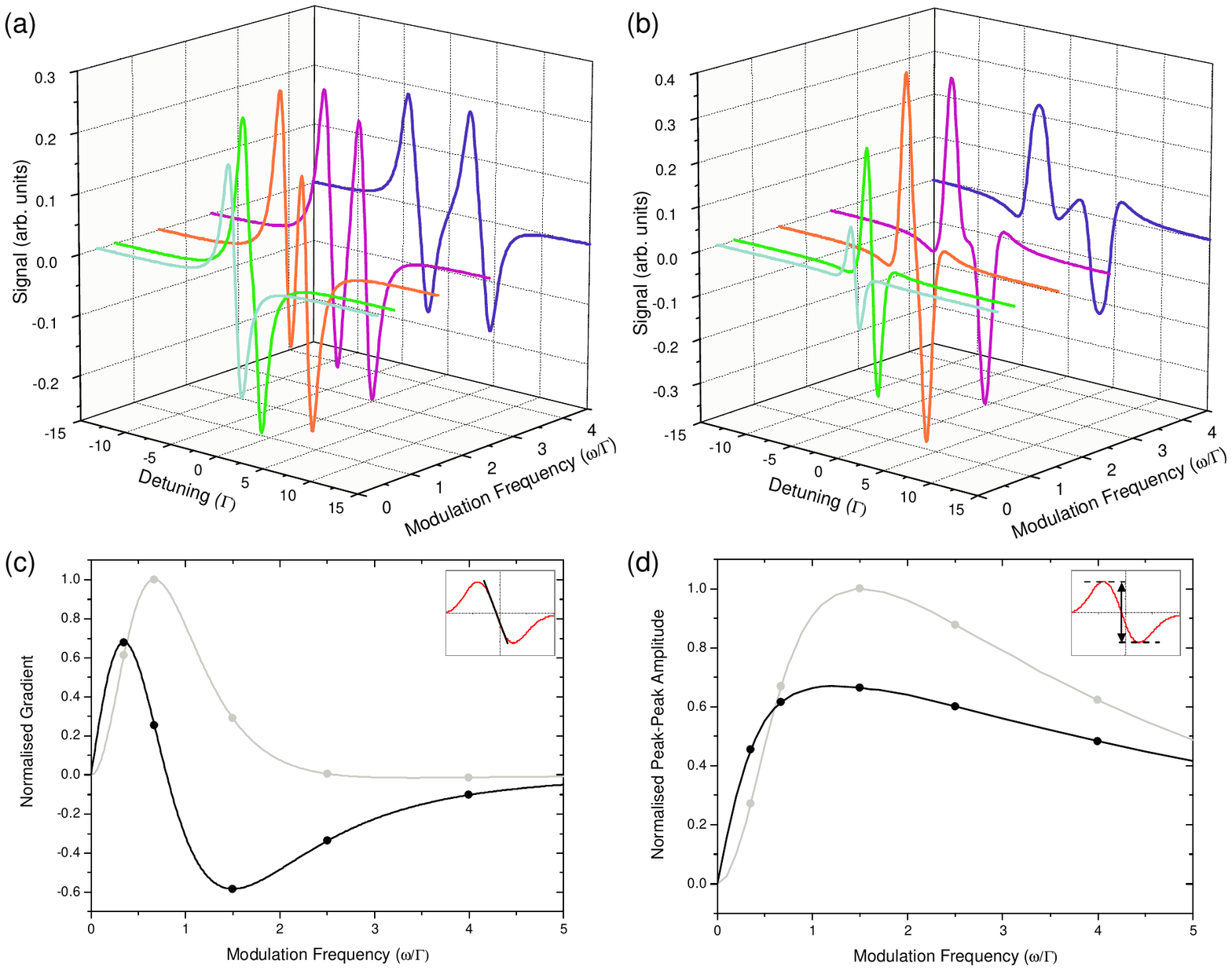}
\caption{Theoretical modulation transfer signals for the in-phase and quadrature components are shown in (a) and (b) respectively. Modulation frequencies of $\omega_{m}/\Gamma$ = 0.35 (light blue), 0.67 (green), 1.50 (orange), 2.50 (purple) and 4.00 (blue) are displayed in both plots. These data were calculated using the cosine and sine terms of equation \ref{eq:S1} respectively. (c) and (d) show normalised signal gradients of the zero crossing and peak to peak amplitudes respectively, as a function of modulation frequency for the in-phase (black) and quadrature (grey) components of the signal. The points in (c) and (d) mark the theoretical signals shown in (a) and (b).}\label{fig:theory}
\end{figure}

Modulation transfer spectroscopy is a pump-probe technique which produces sub-Doppler lineshapes suitable for laser locking \cite{Bertinetto01}. In this paper we will refer to the two counter-propagating laser beams as the pump and probe beams as in standard saturation absorption/hyperfine pumping spectroscopy, however it should be noted that the beam powers are approximately equal. The underlying principle of modulation transfer spectroscopy is as follows. An intense, single frequency pump beam is passed through an EOM, driven by an oscillator at frequency $\omega_{m}$. The transmitted phase-modulated light can be represented in terms of a carrier frequency $\omega_{c}$ and sidebands separated by the modulation frequency $\omega_{m}$
\begin{displaymath}
E=E_{0}\rm{sin}[\omega_{c}t + \delta\rm{sin}\omega_{m}t],
\end{displaymath}
\begin{equation}
E=E_{0}\Bigg[\sum_{n=0}^{\infty} J_{n}(\delta){\rm{sin}}(\omega_{c}+n\omega_{m})t+\sum_{n=1}^{\infty}(-1)^{n} J_{n}(\delta){\rm{sin}}(\omega_{c}-n\omega_{m})t\Bigg],
\end{equation}
\noindent where $\delta$ is the modulation index and $J_{n}(\delta)$ is the Bessel function of order $n$. Usually the modulation index $\delta<1$, so that the probe beam can be adequately described by a strong carrier wave at $\omega_{c}$ and two weak sidebands at $\omega_{c}\pm \omega_{m}$. The phase modulated pump beam and the counter-propagating, unmodulated probe beam are aligned collinearly through a vapour cell. If the interactions of the pump and probe beams with the atomic vapour are sufficiently nonlinear, modulation appears on the unmodulated probe beam. This modulation transfer has been described as an example of four-wave mixing \cite{Raj80,Ducloy82}. Here, two frequency components of the pump beam combine with the counter propagating probe beam by means of the non-linearity of the absorber (the third order susceptibility, $\chi^{(3)}$), to generate a fourth wave - a sideband for the probe beam. This process occurs for each sideband of the pump beam. The strongest modulation transfer signals are observed for closed transitions; here four-wave mixing is a very efficient process as atoms cannot relax into other ground states. Modulation transfer only takes place when the sub-Doppler resonance condition is satisfied and, in this way, the lineshape baseline stability becomes almost independent of the residual linear-absorption effect. This leads to a flat, zero background signal and is one of the major advantages of modulation transfer spectroscopy. The stability of the lineshape baseline is therefore independent of changes in absorption due to fluctuations in polarisation, temperature and beam intensity. Another advantage is that the position of the zero-crossing always falls on the centre of the sub-Doppler resonance, and is not effected by, for example, magnetic field or wave plate angle-dependent shifts which upset both DAVLL and polarisation spectroscopy. After passing through the vapour cell the probe beam is incident on a photo-detector. The probe sidebands generated in the vapour beat with the probe beam to produce alternating signals at the modulation frequency $\omega_{m}$. The beat signal on the detector is of the form \cite{Shirley82}

\begin{displaymath}
S(\omega_{m})=\frac{C}{\sqrt{\Gamma^{2}+\omega_{m}^{2}}}\sum_{n=-\infty}^{\infty}J_{n}(\delta)J_{n-1}(\delta)
\end{displaymath}
\begin{displaymath}
\   \times[(L_{(n+1)/2}+L_{(n-2)/2}){\rm{cos}}(\omega_{m}t+\phi)
\end{displaymath}
\begin{equation} \label{eq:S}
\   +(D_{(n+1)/2}+D_{(n-2)/2}){\rm{sin}}(\omega_{m}t+\phi)],
\end{equation}
where
\begin{equation}
\centering
L_{n}=\frac{\Gamma^{2}}{\Gamma^{2}+(\Delta-n\omega_{m})^{2}},
\end{equation}
and
\begin{equation}
\centering
D_{n}=\frac{\Gamma(\Delta-n\omega_{m})}{\Gamma^{2}+(\Delta-n\omega_{m})^{2}}.
\end{equation}
\noindent Here $\Gamma$ is the natural linewidth, $\Delta$ is the frequency detuning from line centre and $\phi$ is the detector phase with respect to the modulation field applied to the pump laser. The constant $C$ represents all the other properties of the medium and the probe beam that are independent of the parameters described above. If we assume that $\delta<1$ and consider only the first order sidebands then equation \ref{eq:S} is simplified to
\begin{displaymath}
S(\omega_{m})=\frac{C}{\sqrt{\Gamma^{2}+\omega_{m}^{2}}}J_{0}(\delta)J_{1}(\delta)
\end{displaymath}
\begin{displaymath}
\times[(L_{-1}-L_{-1/2}+L_{1/2}-L_{1}){\rm{cos}}(\omega_{m}t+\phi)
\end{displaymath}
\begin{equation} \label{eq:S1}
+(D_{1}-D_{1/2}-D_{-1/2}+D_{-1}){\rm{sin}}(\omega_{m}t+\phi)].
\end{equation}
In equations \ref{eq:S} and \ref{eq:S1} the sine term represents the quadrature component of the signal and the cosine term the in-phase component of the signal. Using a phase-sensitive detection scheme, it is therefore possible to recover the absorption and dispersion components of the sub-Doppler resonance by setting the phase of the reference signal to select either the quadrature or in-phase signal component, respectively \cite{Shirley82}. Theoretical absorption and dispersion signals are shown in figures \ref{fig:theory} (a) and (b), respectively. Both figures show signals at modulation frequencies of $\omega_{m}/\Gamma$ = 0.35, 0.67, 1.50, 2.50 and 4.50. These plots show that both signals are odd functions of the detuning between the laser frequency and the resonance frequency, and provided that $\omega_{m} \le \Gamma$, both the absorption and dispersion signals have a similar line-shape with a large gradient when crossing the centre of the resonance. This makes them ideal candidates to be used as `error signals' for laser locking. It should be noted that when $\omega_{m} \le \Gamma$ it is difficult to distinguish between the in-phase and quadrature components of the signal as they both have a dispersive-like lineshape. The normalised zero crossing gradient and peak to peak values for both signal types are shown as a function of modulation frequency in figures \ref{fig:theory} (c) and (d) respectively. These plots show that the absorption signals (the in-phase component) have a maximum gradient when $\omega_{m}/\Gamma$ = 0.35 and a maximum peak to peak amplitude when $\omega_{m}/\Gamma$ = 1.20; dispersion signals (the quadrature component) have a maximum gradient when $\omega_{m}/\Gamma$ = 0.67 and a maximum peak to peak amplitude when $\omega_{m}/\Gamma$ = 1.50. The maximum gradient of the absorption signal is $0.68$ times the maximum gradient of the dispersion signal and the maximum peak to peak value for the absorption signal is $0.67$ times the maximum peak to peak value of the dispersion signal. For these data the values of $C$, $J_{0}(\delta)$, and $J_{1}(\delta)$ were set to unity.
For equations \ref{eq:S} and \ref{eq:S1} it is known that, typically, the maximum signals obtained at any particular modulation frequency do not occur when the detector phase is set solely for the in-phase component or the quadrature component. In general, a mix of the two components is required to produce the maximum signal \cite{Jaatinen95}.

\section{Experiment}

\begin{figure}[t]
\centering
\includegraphics[angle=270, trim = 10mm 10mm 100mm 90mm, clip, scale=0.8]{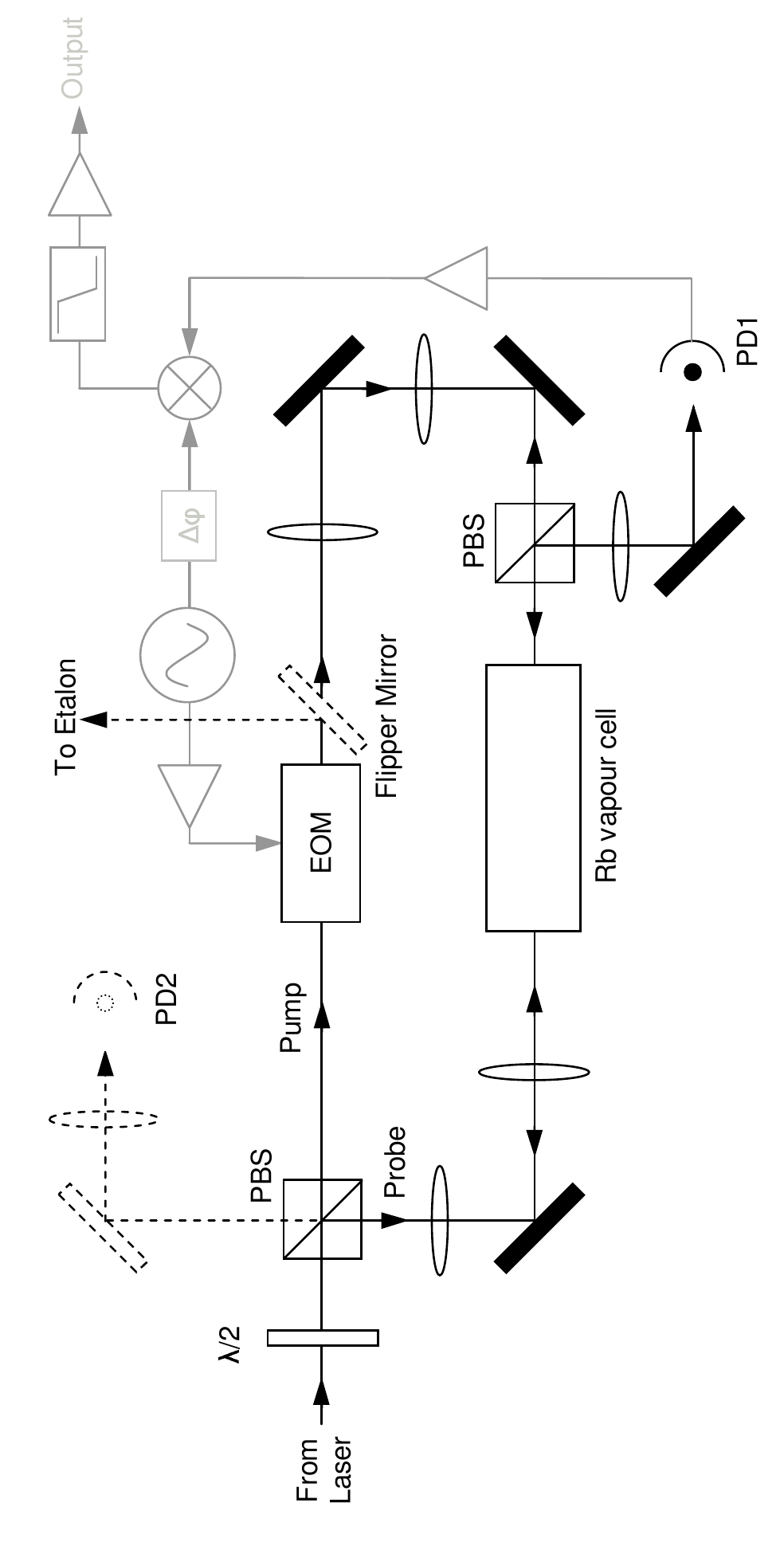}
\caption{Schematic diagram of the modulation transfer spectroscopy
experimental setup, (PBS = polarizing beamsplitter, PD = photodiode, EOM = electro-optic modulator) including the four lens' used to expand the pump and probe beams. The dashed beam path to the second photodiode (PD2) can be used to observe the FM spectroscopy signal using the same setup. To monitor the sidebands applied to the pump beam a flipper mirror was used to send the modulated light to an etalon.}
\label{fig:setup}
\end{figure}

The experimental set-up is shown in figure \ref{fig:setup}.
The experiment used a grating stabilized Toptica DL100 diode laser to provide light at 780~nm with an optical isolator to prevent light from
being reflected back into the laser. The optical setup utilized two narrow-band polarizing beam splitters (Casix PBS0101) to split and then overlap the light. A low-order half-wave plate (Casix WPL1210) controlled the power ratio of the pump and probe beams. For the majority of the measurements $(2.90\pm0.05)$ mW of light was available. The half wave-plate was set to give the maximum signal amplitude which corresponded to a probe laser power of $(1.55\pm0.05)$ mW. Telescopes consisting of 25 mm and 100 mm lens' were used to increase the beam size of both the probe and pump beams by a factor of approximately four. Without the telescopes the pump and probe beams had mean $1/e^{2}$ radii of $(0.54\pm0.01)$ mm and $(0.52\pm0.01)$ mm respectively. The probe beam was aligned collinearly with the counter-propagating, pump beam through a 5 cm long room temperature vapour cell and then detected on a fast photodiode (EOT ET-2030A) with a responsivity of 500 V/W. The signal from the photodiode was amplified (Mini-Circuits ZFL-500LN) before reaching a frequency mixer (Mini-Circuits ZAD-6+). The output signal from the mixer was amplified by a factor of 200, through a 10 kHz low pass active filter, which also served to remove the high frequency signal from the mixer. The phase of the reference signal supplied to the mixer was changed by altering the cable length between the oscillator (SRS DS345 function generator) and the mixer. At each modulation frequency the phase of the signal was optimized to give the maximum peak to peak amplitude of the output signal. This did not occur when the detector phase was set solely for the in-phase or the quadrature component of the signal, but when the detector phase was a mix of the two components \cite{Jaatinen95}.


\begin{figure}[t]
\centering
\includegraphics[angle=270, trim = 10mm 10mm 45mm 90mm, clip, scale=0.5]{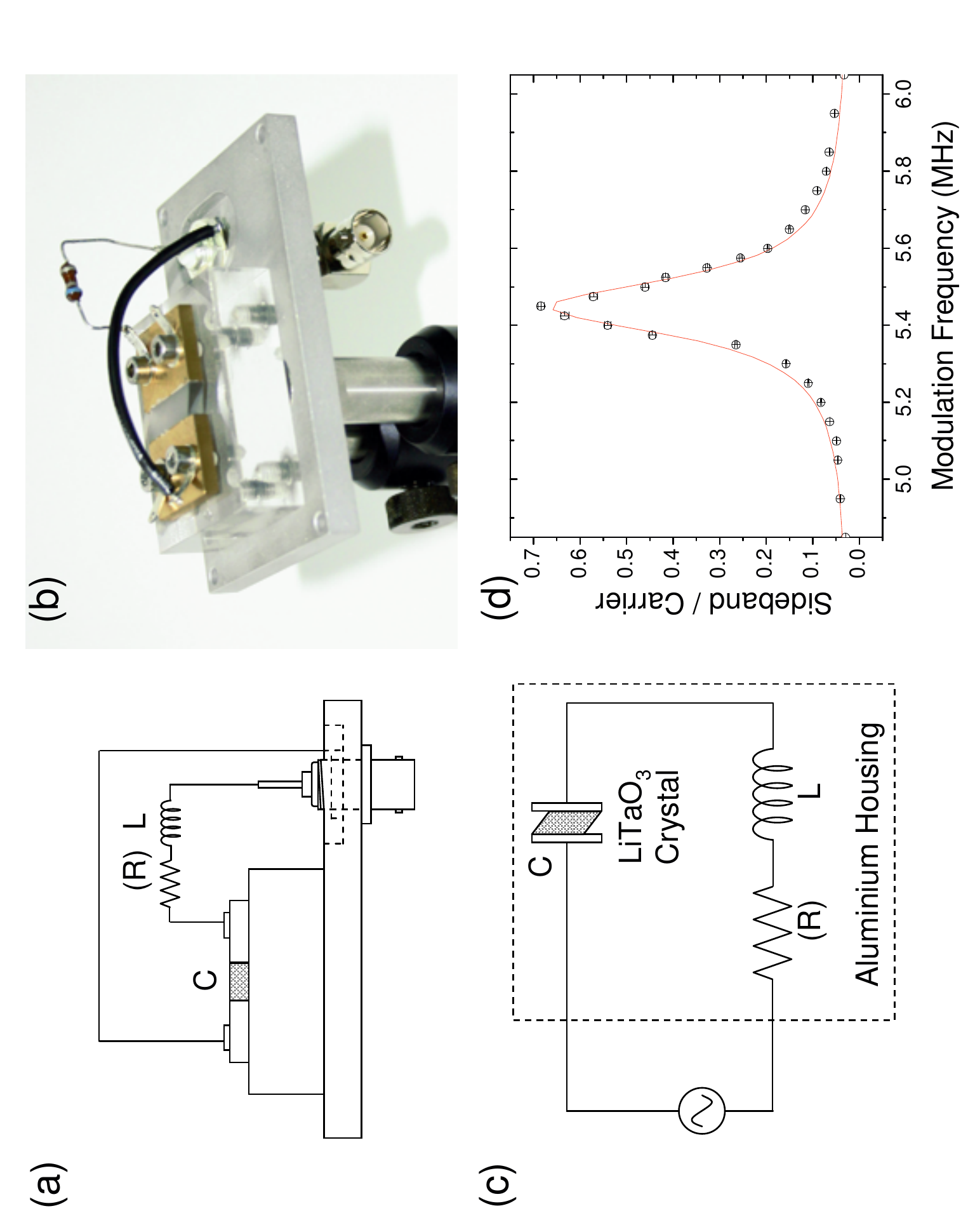}
\caption{(a) Schematic diagram of the EOM setup with (b) a
photograph of the same setup. (c) A simple LCR circuit where L is
the inductance of the inductor, R is the resistance of the inductor and C is the capacitance of the
crystal. (d) By plotting the sideband height as a fraction of
the carrier height it is possible to map out the resonance as a
function of frequency, a Lorentzian fit gives a Q-Factor of 15(1).}\label{fig:EOM}
\end{figure}

The home-built EOM assembly can be seen in figures \ref{fig:EOM} (a) and (b). The device uses a Brewster-cut lithium tantalate (LiTaO$_{3}$) crystal (supplied by Leysop Ltd.). The crystal measures 6 mm wide, 3 mm deep and 17 mm long. The Brewster angled faces of the crystal were optically polished to laser finish and the sides (Z-faces) were coated with chrome-gold to form electrodes. The crystal was mounted between two brass electrodes, using silver paint to ensure good electrical contact, so that an electric field could be applied transverse to the optical axis. The crystal and electrodes were then mounted on a perspex block within an aluminium housing. The crystal was connected in a simple resonant LCR circuit (see figure \ref{fig:EOM} (c)) driven by the amplified (Mini-Circuits ZHL-3A) output from the oscillator in order to enhance the amplitude of the ac voltage across the crystal. The resonant frequency of the circuit, $\omega$, is  $\omega=\frac{1}{\sqrt{LC}}$ \cite{Grant}, where $L$ is the inductance of the inductor and $C$ is the parallel plate capacitance of the crystal between the two electrodes. Figure \ref{fig:EOM} (d) shows a typical resonance at 5.45 MHz. On resonance the voltage across the crystal is increased with respect to the voltage applied to the circuit by a factor equal to the Q-factor of the circuit. By fitting a Lorentzian curve to the plot in figure \ref{fig:EOM} (d) data the Q-factor of the circuit was found to be $(15\pm 1)$.

Modulation transfer spectroscopy signals were recorded both with and without the telescopes for resonant modulation frequencies of 5.45, 7.20, 10.50, 12.35, 14.90 and 19.80 MHz for the $^{87}$Rb $F=2 \rightarrow F^{'}=3$ and $^{85}$Rb $F=3 \rightarrow F^{'}=4$ transitions. For frequency calibration we employed a separate saturated absorption/hyperfine pumping spectroscopy set-up \cite{MacAdam92,Smith04}. The resonant frequency was varied by changing the value of the inductor, $L$, in series with the crystal from a value 100 $\mu$H for the 5.45 MHz resonance to 15 $\mu$H for the 19.80 MHz resonance. All of the inductors used in this investigation were high frequency / RF type inductors. As shown in figure \ref{fig:setup} a flipper mirror was mounted after the EOM. When in position this sent light into an etalon with a 300 MHz free spectral range (Coherent, 33-6230-001) and allowed the sideband/carrier ratio to be measured. During the experiment this ratio was kept at a constant value of $(13.2 \pm 0.2)\ \%$ for all measurements to be in the regime of having only first order sidebands. This was controlled by changing the peak to peak amplitude of the signal from the oscillator over the range from 0.85 V to 1.40 V.

By positioning a second fast photodiode (EOT ET-2030A) at the position marked by PD2 in figure \ref{fig:setup} it was possible to record both modulation transfer spectroscopy signals and FM spectroscopy signals simultaneously. This allowed for a direct comparison between the two methods to be made. These measurements were recorded without telescopes in the setup at a modulation frequency of 12.35 MHz. For these data the probe beam power was $(498\pm 5)$ $\mu$W and the pump beam power was $(814\pm 5)$ $\mu$W.

\section{Results and Discussion}\label{results}

\subsection{Comparing modulation transfer spectroscopy with FM spectroscopy}
\begin{figure}[t]
\begin{center}
\includegraphics[angle=270, trim = 0mm 10mm 8mm 10mm, clip, scale=0.4]{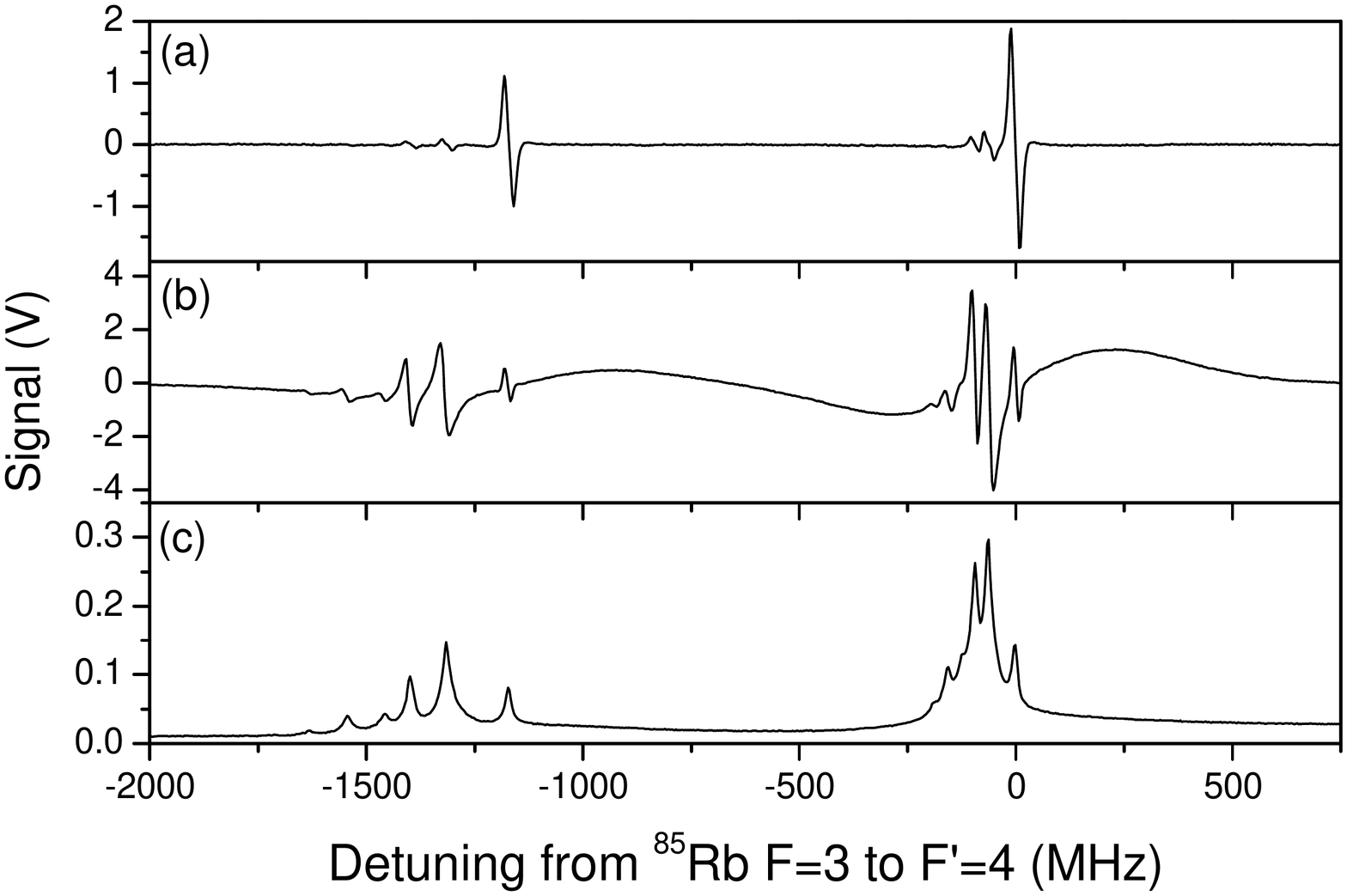}
\end{center}
\caption{\footnotesize{A comparison between (a) modulation transfer
spectroscopy and (b) FM spectroscopy for the $^{87}$Rb $F=2
\rightarrow F^{'}$ and $^{85}$Rb $F=3 \rightarrow F^{'}$ transitions at a modulation frequency of 12.35 MHz. (c) The reference saturated absorption/hyperfine pumping spectroscopy signal is included for completeness. A 10 point moving average has
been applied to the data.}}\label{fig:comparison}
\end{figure}

Figures \ref{fig:comparison} (a), (b) and (c) show a modulation transfer signal, an FM signal and a saturated absorption/hyperfine pumping spectroscopy signal respectively for the $^{87}$Rb $F=2 \rightarrow F^{'}$ and $^{85}$Rb $F=3
\rightarrow F^{'}$ transitions in rubidium. These figures highlight the differences between modulation transfer spectroscopy and FM spectroscopy.
The modulation transfer signal has a very flat, zero background signal. This is due to modulation transfer only taking place when the sub-Doppler resonance condition is satisfied; hence the baseline stability is almost independent of the residual linear-absorption effect. In contrast to this, the FM signal is observed on a sloping background, approximating to the derivative of the Doppler-broadened absorption profile. Usually, a second stage of demodulation is employed in FM spectroscopy to remove this background by amplitude modulating the pump beam. This extra stage of complexity  (and cost) is not needed for modulation transfer spectroscopy. The modulation transfer signal is always dominated by one zero crossing with a large peak to peak amplitude for the transitions from each hyperfine ground state. The signal with the biggest peak to peak amplitude always corresponds to the closed transition. This can be advantageous, particularly in the case where there are many closely spaced hyperfine transitions, such as the $^{85}$Rb $F=2 \rightarrow F^{'}$ transitions. In contrast, the FM signal displays the same number of lines as the standard saturated absorption/hyperfine pumping spectroscopy signal. Whilst this can be an advantage in applications where one wants to lock away from the closed transition, it can also lead to confusion when locking to a particular line in a closely spaced spectrum.



\subsection{Dependence of the modulation transfer signal on beam size and modulation frequency}
\begin{figure}[h!]
\centering
\includegraphics[angle=270, trim = 15mm 0mm 105mm 80mm, clip, scale=0.8]{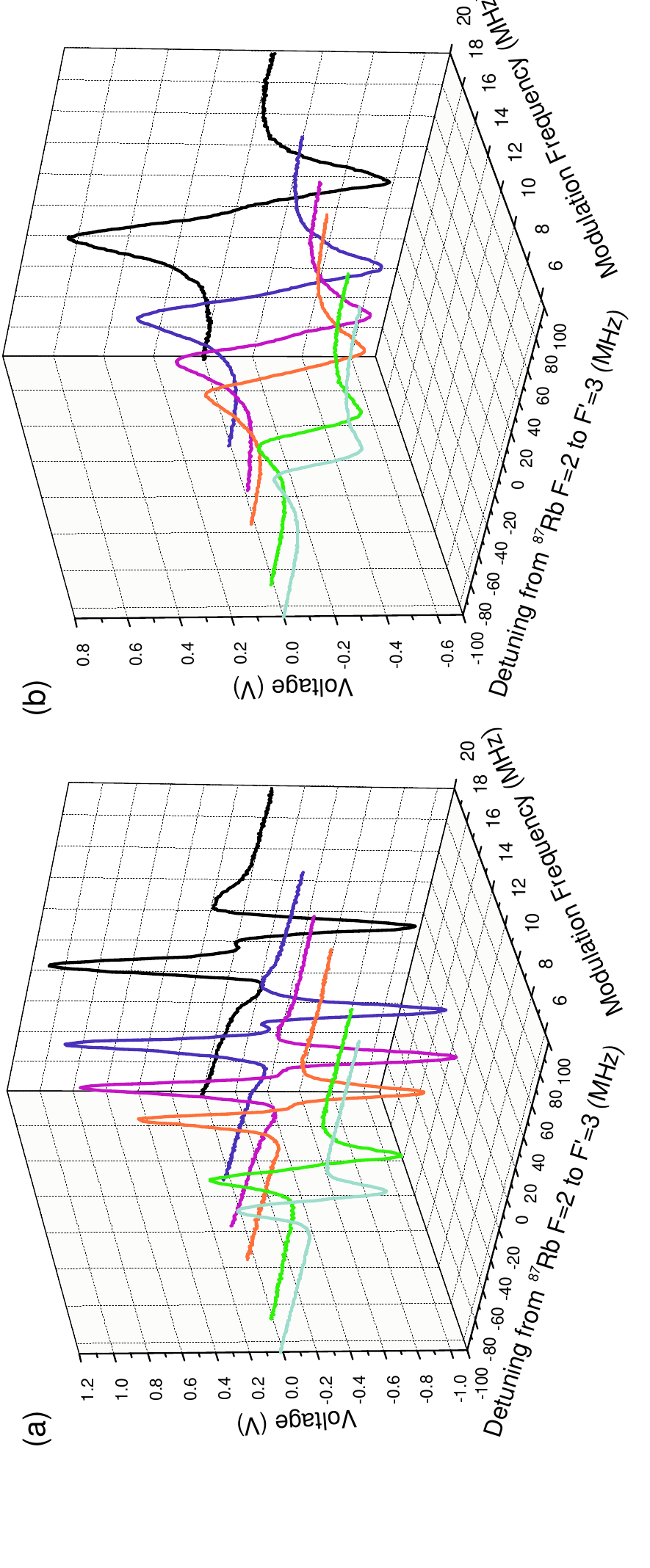}
\caption{Experimental modulation transfer spectroscopy signals on the $^{87}$Rb $F=2
\rightarrow F^{'}=3$ transition as a function of modulation frequency
with telescopes in the setup (a), and with no telescopes in the setup (b). The
modulation frequencies are 5.45 MHz (light blue), 7.20 MHz (green), 10.50
MHz (orange), 12.35 MHz (purple), 14.90 MHz (blue) and 19.80 MHz
(black). The data were obtained with a sideband/carrier ratio of $
(13.2 \pm 0.2)\%$. A 10 point moving average has
been applied to the data.}\label{fig:lineshape}
\end{figure}

Figure \ref{fig:lineshape} shows data for the $^{87}$Rb $F=2
\rightarrow F^{'}=3$ transition with (a), and without (b), the telescopes in position, as shown in figure \ref{fig:setup}. With the telescopes in position the pump and probe beams had mean $1/e^{2}$ radii of $(2.18\pm0.04)$ mm and $(2.13\pm0.07)$ mm respectively.  Modulation frequencies of 5.45, 7.20, 10.50, 12.35, 14.90 and 19.80 MHz were investigated for both cases.
These figures clearly show that, for the same beam powers, including telescopes in the setup to expand the beams improves the signal. This improvement follows from the narrower effective sub-Doppler line-width, $\Gamma'$, in the case where the beam intensity is reduced, which then results in sharper modulation transfer signals. It is important to note that as the signal also increases with the total amount of power in the beams, this improvement cannot be achieved by simply reducing the total power in the setup. Figure \ref{fig:lineshape} (a) shows that as the modulation frequency, $\omega_{m}$, becomes greater than the sub-Doppler line-width $\Gamma$, a `kink' appears in the locking slope. If the modulation frequency is increased further this `kink' becomes more pronounced until eventually the zero crossing gradient changes sign (cf. figures \ref{fig:theory} (a) and (b)). Generally the signals shown in figure \ref{fig:lineshape} (b), for the case without the telescopes, do not show this behavior as the effective sub-Doppler line-width, $\Gamma'$, has been increased due to power broadening. Despite this broadening, a small decrease in the zero crossing gradient can be seen for the 19.80 MHz signal as again the modulation frequency $\omega_{m}$, becomes greater than the effective sub-Doppler line-width $\Gamma'$.
\begin{figure}[t]
\begin{center}
\includegraphics[angle=270, trim = 10mm 10mm 30mm 75mm, clip, scale=0.6]{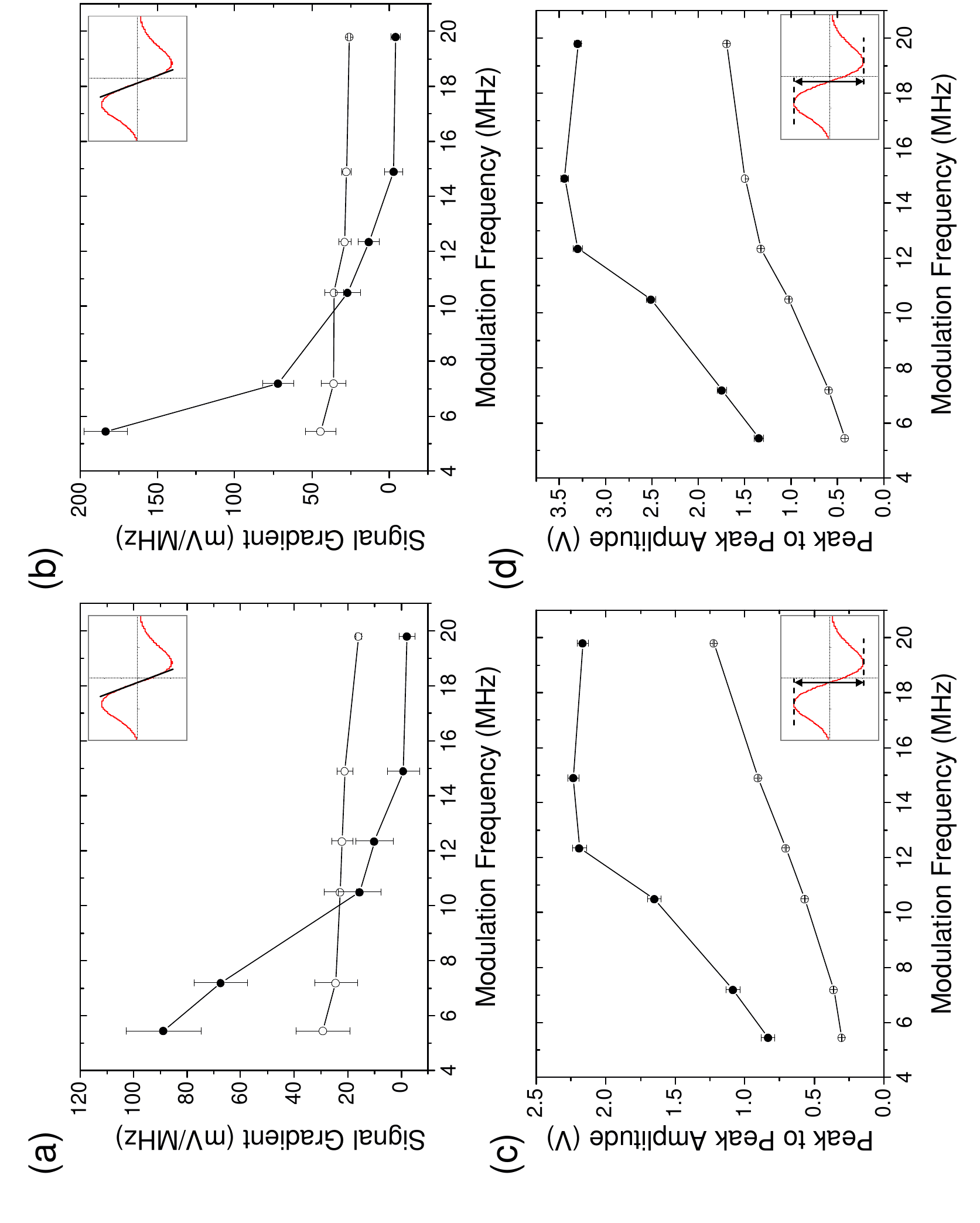}
\end{center}
\caption{\footnotesize{Gradients of the zero crossings of the
modulation transfer signal as a function of modulation frequency for
(a) the $^{87}$Rb $F=2 \rightarrow F^{'}=3$ and (b) the $^{85}$Rb $F=3
\rightarrow F^{'}=4$ transitions. Peak to Peak values of the
modulation transfer signal as a function of the modulation frequency
for (c) the $^{87}$Rb $F=2 \rightarrow F^{'}=3$ and (d) the $^{85}$Rb
$F=3 \rightarrow F^{'}=4$ transitions. Data are shown without telescopes in
the setup (open symbols) and with telescopes in the setup (solid
symbols).}}\label{fig:comparingGRAD&PK}
\end{figure}
The data of figure \ref{fig:comparingGRAD&PK} show the evolution of the signal gradient and amplitude for the $^{87}$Rb $F=2 \rightarrow F^{'}=3$ transition, (a) and (c) respectively, and the $^{85}$Rb $F=3
\rightarrow F^{'}=4$ transition, (b) and (d) respectively, for different modulation frequencies. Data sets were taken with telescopes in the setup (closed symbols), and without telescopes in the setup (open symbols). The data show that without telescopes in the setup, as the modulation frequency is increased, the signal gradient decreases and the signal amplitude increases monotonically. However with telescopes in the setup, as the modulation frequency increases the signal gradient decreases and reaches approximately zero for a modulation frequency of 14.90 MHz for both the $^{87}$Rb $F=2 \rightarrow F^{'}=3$ and the $^{85}$Rb $F=3\rightarrow F^{'}=4$ transitions. The data show that the signal peak to peak amplitude increases to a maximum value and then begins to decrease. This maximum peak to peak value occurs for a modulation frequency of around 14 MHz for both transitions.

The experimental data is in good qualitative agreement with the predictions of the theoretical lineshape model discussed in section \ref{theory}. However, the quantitative analysis does not show such good agreement. For example when fitting the theory for dispersion line-shapes to the experimental data recorded with telescopes in the setup, we obtain an effective linewidth of ($6.5\pm0.5$) MHz from the gradient data and ($12.0\pm0.5$) MHz from the amplitude data. This behavior is explained by the fact that in setting the phase to give the maximum signal amplitude we are in fact observing a mix of the in-phase and quadrature components. Using a more sophisticated model consisting of the mix of in-phase and quadrature components to give the maximum peak to peak amplitude we observe better, but still not perfect, agreement with the data. This analysis gives the reasonable effective sub-Doppler linewidths of around 9 MHz and 14 MHz with and without telescopes in the setup, respectively. These linewidths are greater than the natural linewidth of $\Gamma =2\pi\times6.065(9)$ MHz \cite{Smith04}, most probably due to power broadening. However, the intensity dependence of the effective sub-Doppler linewidth in modulation transfer spectroscopy is not trivial to calculate because the underlying four-wave mixing process cannot be treated as a simple two level system. For example, for the low modulation index used in this work, $\delta$, the intensity of the sidebands involved in the four-wave mixing process is greatly reduced compared to the carrier intensity. Additionally four-wave mixing only occurs at detunings of $\omega_{0}\pm\omega_{m}/2$ and $\omega_{0}\pm\omega_{m}$, where $\omega_{0}$ is the frequency of the sub-Doppler resonance. This complexity becomes apparent if the effective sub-Doppler linewidth is calculated using \cite{Loudon}
\begin{equation}
\Gamma' = \Gamma \sqrt{1+\frac{I}{I_{SAT}}},\label{eq:broad}
\end{equation}
where $I$ is the beam intensity and $I_{SAT}$ is the saturation intensity for linearly polarized light on the D2 transition in rubidium. For the intensities used in this experiment, equation \ref{eq:broad} gives effective sub-Doppler linewidths of approximately $3\Gamma$ ($\approx 18\,\mbox{MHz})$ and $12\Gamma$ $(\approx 72\,\mbox{MHz})$ with and without telescopes in the setup respectively. Clearly these values are in strong disagreement with our observations and indicate that large intensities can be used for modulation transfer spectroscopy without increasing the effective sub-Doppler linewidth. The relatively high intensities used in this investigation are in part an artifact of the low gain of the fast photodiode used. We should note that another modulation transfer setup in Durham uses a photodiode with much higher gain and observes comparable signals to those presented here with beam intensities of the order of $I_{SAT}$. Interestingly, these signals still show the dominance of closed transitions as this is an artifact of the four-wave mixing process.

\subsection{A modulation transfer signal for the $^{85}$Rb repump transition}\label{repump}

\begin{figure}[t]
\centering
\includegraphics[angle=270, trim = 0mm 10mm 10mm 10mm, clip, scale=0.4]{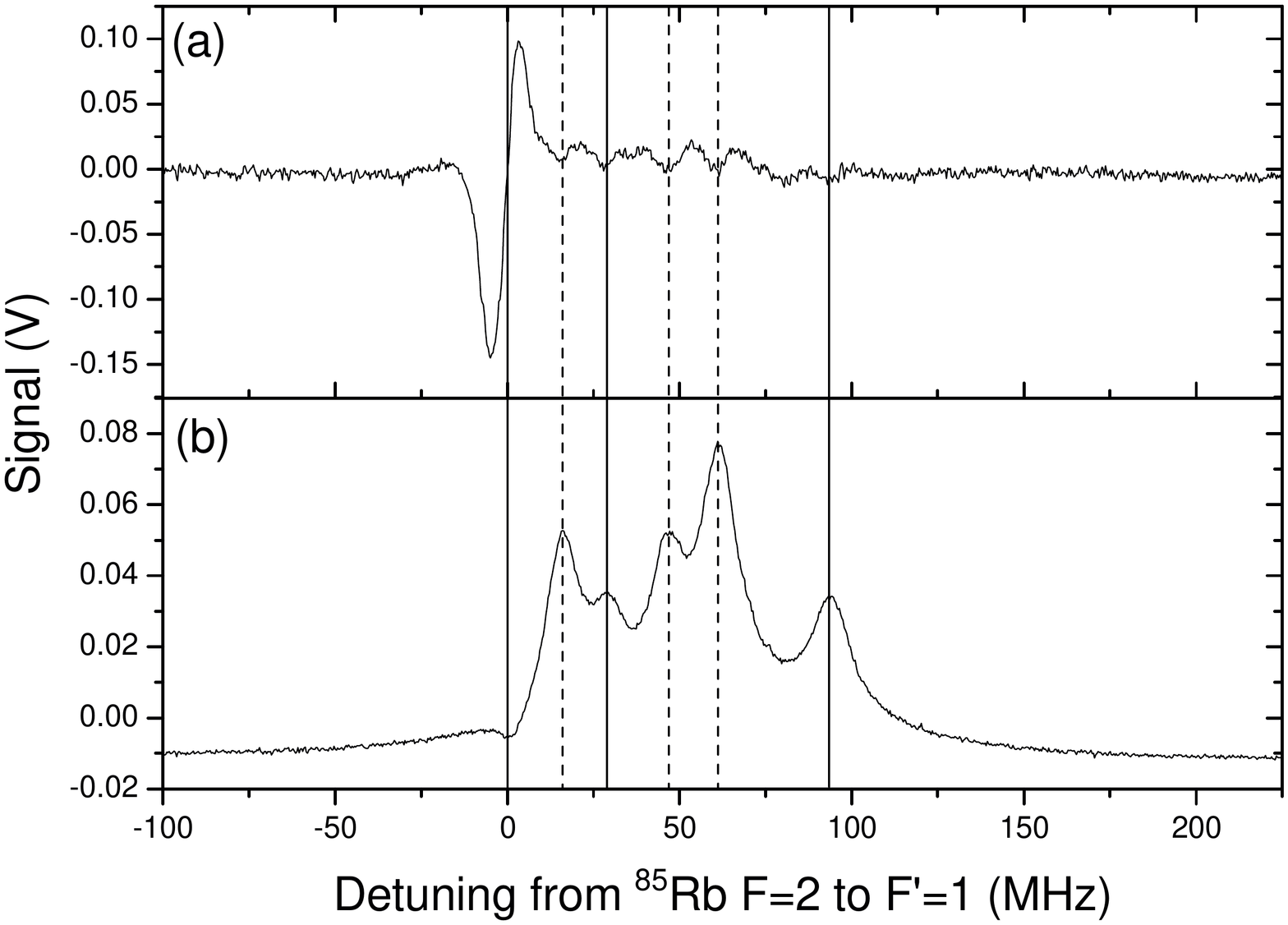}
\caption{(a) Modulation transfer spectroscopy signal for the $^{85}$Rb $F=2
\rightarrow F^{'}$ transitions at a modualtion frequency of 7.20 MHz. (b) Pump-probe spectroscopy signal for the $^{85}$Rb $F=2
\rightarrow F^{'}$ transitions, solid lines mark hyperfine transitions and dashed lines mark crossovers. A 10 point moving average has been applied to the data.} \label{fig:repump}
\end{figure}

The modulation transfer signal recorded for the $^{85}$Rb $F=2
\rightarrow F^{'}$ transitions, at a modulation frequency of 7.20 MHz, is shown in figure \ref{fig:repump} (a), with the corresponding saturated absorption/hyperfine pumping spectroscopy signal shown in figure \ref{fig:repump} (b). For the $^{85}$Rb $F=2\rightarrow F^{'}$ transitions the excited state hyperfine splitting is spread over just $\approx 93$ MHz, this leads to closely spaced sub-Doppler features in pump-probe spectroscopy that are difficult to resolve, as shown in figure \ref{fig:repump} (b). Note that the $^{85}$Rb $F=2
\rightarrow F^{'}=1$ transition shown in this figure appears as a dip rather than a peak due to optical pumping. However, as modulation transfer spectroscopy signals are dominated by closed transitions, this method generates a clear, unambiguous frequency discriminant even when there are many closely spaced sub-Doppler features, as shown in figure \ref{fig:repump} (a). These data were recorded with the telescopes in the setup. To optimize the signal the probe beam power was set to $(2.71\pm0.05)$ mW and the power in the pump beam was changed to $(0.98\pm0.05)$ mW. We note that acousto-optic modulators (AOMs) would be required to provide a small frequency detuning in the case of the cooling transition and to bridge the gap from the $F=2 \rightarrow F^{'}=1$ transition to the repump transition ($F=2 \rightarrow F^{'}=3)$.

\section{Conclusions}
In summary, we have characterized modulation transfer spectroscopy for the rubidium $D2$ transitions and have highlighted the advantages of this technique for laser frequency stabilization. These advantages include a flat zero background signal, and the signal being dominated by closed transitions leading to one zero crossing with a large peak to peak amplitude for each hyperfine transition. We have demonstrated this useful property for the $^{85}$Rb $F=2 \rightarrow F^{'}$ transitions where the excited state hyperfine splitting is spread over just $\approx 93$ MHz, and the closely spaced features in standard pump-probe spectroscopy are difficult to resolve. We have shown that both the modulation transfer signal gradient and peak to peak amplitude can be significantly improved through a careful choice of $\omega_{m}/\Gamma$ and by expanding the beams using telescopes. Despite a long history modulation transfer spectroscopy is not widely discussed in the literature. However, we hope this paper demonstrates that the technique has many benefits and is well-suited to applications in laser cooling and trapping of alkali gases.


\ack This work was funded by the Engineering and Physical Sciences Research Council (grant EP/E041604/1) and the European Science Foundation within the EUROCORES Cold Quantum Matter (EuroQUAM) programme. SLC acknowledges the support of the Royal Society. The authors acknowledge many fruitful discussion with Hanns-Christoph N\"{a}gerl and thank M.P.A. Jones for a careful reading of the manuscript.

\section*{References}

\bibliography{modtransferv3}


\end{document}